\documentclass[10pt,twocolumn]{article}

\usepackage[a4paper,margin=0.75in]{geometry}
\usepackage[T1]{fontenc}
\usepackage[utf8]{inputenc}
\usepackage{lmodern}

\usepackage{graphicx}
\usepackage{booktabs}
\usepackage{tabularx}
\usepackage{url}
\usepackage[hidelinks]{hyperref}
\usepackage{authblk}
\usepackage{orcidlink}
\usepackage{abstract}
\usepackage{balance}

\usepackage{caption}
\usepackage[backend=biber,style=numeric,doi=true,url=false,isbn=false,sortcites=true,giveninits=true]{biblatex}
\title{An Exploratory Case Study of LLM-Assisted Refactoring and Gameplay Feature Generation in an Endless Runner Game}

\author[1]{Jan Wunderlich\orcidlink{0009-0004-8909-8653}}
\author[1]{Markus Kleffmann\orcidlink{0009-0004-4175-552X}}
\author[1]{Sebastian Lempert\orcidlink{0000-0002-5214-5944}}
\affil[1]{
    IU International University of Applied Sciences, Erfurt, Germany\\
    \href{mailto:jan.wunderlich@iu-study.org}{jan.wunderlich@iu-study.org}, %
    \href{mailto:markus.kleffmann@iu.org}{markus.kleffmann@iu.org}, %
    \href{mailto:sebastian.lempert@iu.org}{sebastian.lempert@iu.org}
}

\date{}

\begin{document}

\twocolumn[
\maketitle

\begin{onecolabstract}
Large language models (LLMs) are increasingly used to support software development, but their practical usefulness in applied game-development settings remains underexplored, especially when generated code must be integrated into an existing game software system. This paper presents an exploratory empirical case study of GPT-4o in a custom Python/Pygame endless runner. The study examines six selected development tasks: three localized refactoring tasks and three tasks involving gameplay feature generation. The resulting implementations were evaluated using software metrics, unit tests, and manual gameplay assessments. In this case study, all three selected refactoring tasks were completed successfully in functional terms, whereas only one of the three selected gameplay feature generation tasks resulted in a correctly integrated feature. The findings suggest that, in this setting, GPT-4o handled localized transformations more reliably than tasks requiring new gameplay interactions across multiple existing systems. Given the exploratory single-case design, these results are best interpreted as indicative observations rather than as generalizable evidence of category-level model performance. Overall, the paper contributes a transparent case-based account of the opportunities and limitations of LLM-assisted refactoring and gameplay feature generation in an existing game software system.
\end{onecolabstract}

\noindent\textbf{Keywords:} large language models, GPT-4o, code generation, refactoring, gameplay feature generation, game development, Python, Pygame, case study

\vspace{1em}
]

\section{Introduction}

Large language models (LLMs) are increasingly used to support software development activities such as code generation, code transformation, and interactive programming assistance \cite{ross_programmers_2023, nejjar_llms_2025, liu_guiding_2024}. At the same time, their role in digital games is attracting growing attention, both in the broader research landscape on LLMs and in game-development practice \cite{gallotta_large_2024}. However, despite this growing momentum, there is still limited empirical evidence on how reliably LLMs perform in specific game-development tasks that require integration into an existing game software system rather than isolated benchmark-style code generation \cite{gallotta_large_2024, gu_effectiveness_2025, marini_leveraging_2024}.

This gap is relevant because game development provides a practical context in which generated code must interact with gameplay logic, visual systems, event-driven behavior, and other interdependent runtime components \cite{claypool_latency_2006}. At the same time, many game-development tasks combine structural complexity with recurring patterns, which makes this domain a useful applied setting for studying LLM-supported programming \cite{gu_effectiveness_2025}. Moreover, games allow generated code to be assessed not only through static inspection or unit tests, but also through interactive playtesting, which can reveal behavioral integration problems that may remain hidden in code-level evaluation alone \cite{gallotta_large_2024}.

To investigate this problem, this paper evaluates GPT-4o \cite{openaiGPT4oSystemCard2024} in a concrete game-development setting based on a custom endless runner implemented in Python and Pygame. The study considers two common forms of code-oriented development work within the same existing game codebase: modifying existing structures through refactoring and extending functionality through gameplay feature generation. These task classes are not treated as equivalent in difficulty or scope. Rather, they provide two contrasting perspectives on how GPT-4o behaves when supporting localized code transformations and more integration-intensive feature additions. Accordingly, the study addresses two research questions:

\begin{itemize}
    \item \textbf{RQ1:} Can GPT-4o support the refactoring of an existing video game software system without compromising functionality?
    \item \textbf{RQ2:} Can GPT-4o effectively generate new gameplay features within an existing game architecture?
\end{itemize}

Methodologically, the study comprises two structured experiments with three use cases per task class. The resulting implementations were evaluated using software metrics, unit tests, and manual gameplay assessments.

Overall, this paper contributes an exploratory empirical case study on LLM-assisted game development by examining selected refactoring and gameplay feature generation tasks within the same project context. The study offers a transparent account of where GPT-4o supported code-oriented game-development work in this setting and where integration challenges remained.

\section{Related Work}

Recent research in software engineering has increasingly examined the use of large language models for tasks such as code generation, code transformation, and interactive development assistance \cite{ross_programmers_2023, nejjar_llms_2025, liu_guiding_2024}. At the same time, prior research shows that the effectiveness of LLMs in programming-related tasks depends strongly on factors such as task complexity, prompting strategy, and evaluation context \cite{liu_guiding_2024, liu_refining_2024, liu_no_2024}. Recent empirical work has further shown that LLM-generated code may appear plausible at first glance while still exhibiting defects, maintainability problems, or inconsistent behavior when examined more closely \cite{liu_refining_2024, liu_no_2024, sagodi_methodology_2024, jin_can_2024}.

Within digital games, the use of large language models has recently attracted growing scholarly attention. Review work indicates that LLMs are being explored across a broad range of game-related areas, including game development, game AI, narrative systems, and research-oriented applications \cite{gallotta_large_2024, sweetser_large_2024, yang_gpt_2025_IEEE_TOG}. Beyond this broader landscape, first empirical studies also suggest that LLMs can support concrete technical workflows in game development, for example in failure analysis and debugging-related processes \cite{marini_leveraging_2024}. Taken together, this literature points to a rapidly emerging field with substantial practical interest, but still comparatively limited empirical evidence for specific software-engineering tasks within existing game projects. This is particularly relevant for tasks in which generated or modified code must interact with existing gameplay mechanics, asset handling, input processing, and state-management logic.

Prior work has begun to examine the use of LLMs for refactoring and related code-improvement tasks. Shirafuji et al. \cite{shirafuji_refactoring_2023} showed that LLMs can be used to suggest simpler refactored versions of user-written Python programs, while B. Liu et al. \cite{liu_empirical_2024} provided broader empirical evidence that LLMs can identify refactoring opportunities and recommend useful refactoring solutions, although unsafe or functionality-changing suggestions still occur. Beyond refactoring itself, several studies have highlighted that LLM-generated code often suffers from correctness, maintainability, and security-related issues, even when it appears plausible on the surface \cite{liu_refining_2024, liu_no_2024, siddiq_quality_2024, sagodi_methodology_2024}. Empirical studies of practical usage contexts further suggest that LLM-generated code is often most useful for exploration, examples, or early-stage assistance rather than as immediately production-ready output \cite{jin_can_2024, rasnayaka_empirical_2024}.

Taken together, prior work has examined LLM-assisted refactoring, code generation, and game-related LLM applications from several perspectives. However, these strands are often studied separately. Less is known about how different code-oriented task types behave when they are applied to the same existing game software system and evaluated under the same validation setup. This paper therefore examines refactoring and gameplay feature generation as two contrasting task classes within one shared exploratory case study. The goal is not to treat these task classes as equivalent in difficulty or scope, but to investigate how GPT-4o behaves when supporting localized transformations of existing code compared with more integration-intensive extensions of gameplay functionality.

\section{Methodology}
\label{sec:methodology}

This study follows a structured empirical case-study design to evaluate the use of GPT-4o in a concrete game-development setting. The underlying software artifact is an author-developed endless runner implemented in Python and Pygame, which was chosen as a compact but sufficiently realistic project context for investigating LLM-assisted development tasks.

\begin{figure}[t]
    \centering
    \includegraphics[width=\linewidth]{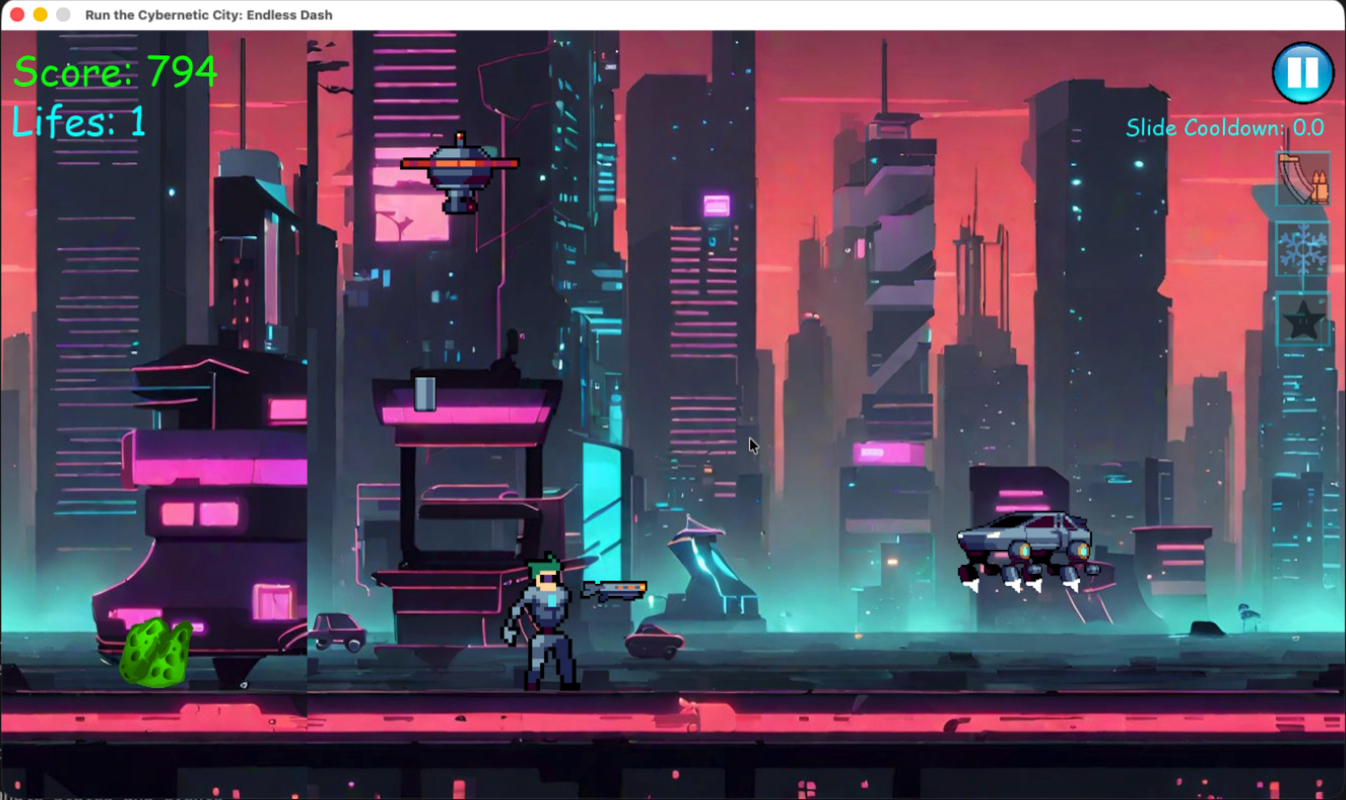}
    \caption{Example gameplay scene from the custom Python/Pygame endless runner used as the software artifact in this study.}
    \label{fig:endless-runner-game}
\end{figure}

In the game, the player controls a character in a continuously scrolling 2D environment and must avoid obstacles, survive enemy interactions, and cover as much distance as possible. The player character can move horizontally, jump, slide, and shoot, while the game includes obstacles such as meteors and cars as well as enemies such as robots and drones. The base game also includes several power-ups, including invincibility, freeze, and weapon enhancements, which affect the player character, enemy behavior, or other gameplay systems. A run ends when the player collides with an obstacle or enemy attack without an active protective effect, making collision handling, player state, and runtime interaction central parts of the game logic. Figure~\ref{fig:endless-runner-game} shows an example scene from the base game, illustrating the player character, enemies, obstacles, and the visual context in which the evaluated code changes were tested.

In structural terms, the project comprised multiple gameplay, menu, asset, and test modules, including central game logic, player-character behavior, enemy behavior, projectile and power-up systems, resource management, and automated pytest-based tests. Across the study, GPT-4o was used in two experiments that focused on refactoring existing code and generating new gameplay features. For each task class, three use cases were selected to cover different levels of complexity while keeping the overall scope manageable. The study therefore examines how GPT-4o behaved across two different forms of code-oriented development work in the same game software system. To support this exploratory comparison, six tasks were selected that represent qualitatively different but practically relevant forms of modification within the same project, while not being pre-calibrated for equivalent difficulty.

The refactoring tasks addressed (R1) the optimization of central state management, (R2) the elimination of redundant asset-loading operations, and (R3) the unification of movement control methods. The state management refactoring (R1) focused on restructuring the central logic responsible for handling game states and events, which previously relied on a strongly nested control flow. The goal was to replace this monolithic decision logic with a more modular event-handling structure that separates state transitions more clearly and reduces overall complexity. The asset-loading refactoring (R2) targeted the part of the code responsible for loading game assets, which previously relied on many repetitive manual image-loading statements. Here, the objective was to replace these duplicated loading operations with a more structured and configurable loading process in order to simplify future changes and reduce redundancy. The movement-control refactoring (R3) addressed previously separate implementations of player-character movement actions such as moving, jumping, and sliding. The corresponding redesign aimed to unify these actions in a more generic control logic so that different movement types could be handled through a single, more compact mechanism.

The tasks involving gameplay feature generation addressed (F1) the implementation of pixel-perfect collision detection, (F2) the introduction of a shrink power-up, and (F3) the implementation of a car platform mechanic. The pixel-perfect collision detection feature (F1) was intended to replace the existing bounding-box-based collision logic with a more precise mask-based approach that considers only visible pixels. This was meant to improve the fairness and realism of collisions in a core gameplay mechanic. The shrink power-up feature (F2) was intended to extend the existing power-up system by temporarily reducing the size of the player character. The goal of this mechanic was to improve evasive capability while preserving correct interaction with other gameplay systems such as jumping, sliding, and weapon handling. The car platform mechanic (F3) was intended to extend the collision logic so that the player character could jump onto moving cars and use them as temporary platforms. This feature therefore introduced a new interaction mechanic that required both specialized collision handling for cars and stable runtime behavior during movement and dismounting.

Taken together, these tasks provided variation in technical scope and integration depth while being situated within the same project context. Because the tasks were not pre-calibrated for equivalent difficulty, this comparison should be interpreted as exploratory and qualitative rather than as a controlled benchmark comparison between refactoring and gameplay feature generation. This task-oriented framing is also consistent with prior empirical work that treats refactoring as a distinct and systematically assessable class of LLM-supported software-engineering activity \cite{cordeiro_empirical_2024, liu_empirical_2024}.

GPT-4o was selected because it was a widely available, high-performing general-purpose LLM with strong code-related capabilities at the time the experiments were conducted. The study therefore evaluates a concrete model snapshot in a specific development setting rather than the current state of all LLM-based coding systems. Because LLM capabilities evolve rapidly, the results should not be interpreted as claims about newer models, specialized code models, or agentic coding tools.

To ensure that GPT-4o had access to the full project context, the complete codebase was provided as a single input rather than being split across multiple prompts. The codebase comprised 31,947 tokens, which was well within the model's available context window and therefore allowed the full project structure to be processed without truncation. At the same time, each refactoring task and each task involving gameplay feature generation was conducted in a separate chat to avoid unintended context carryover between experiments. The complete project is publicly available at GitHub\footnote{\url{https://github.com/jan-wun/PSE_Endless-Runner-Game}}. To further support traceability and reproducibility, each task was also implemented in a dedicated Git branch\footnote{\url{https://github.com/jan-wun/PSE_Endless-Runner-Game/branches/all}} based on the same version of the project's main branch, allowing the code state before and after each task to be inspected at source-code level. Before the actual experiments were conducted, an initial comprehension check was performed by asking GPT-4o to summarize the provided codebase in order to verify that the core structure and dependencies of the project had been correctly interpreted.

The experiments were supported by established prompting techniques selected according to the respective task type. In this study, an iteration denotes one complete interaction cycle with the model, consisting of a prompt, the generated implementation, and its subsequent validation. Iterations were necessary whenever the produced solution did not yet satisfy the expected structural or functional requirements and therefore required refinement through a follow-up prompt. Depending on the use case, the study employed strategies such as self-consistency prompting, few-shot prompting, and test-guided prompting to improve the quality and functional reliability of the generated outputs \cite{sahoo_systematic_2024, knoth_ai_2024, mathews_test-driven_2024}. Table~\ref{tab:prompting-strategies} summarizes which prompting strategies were used for the refactoring and gameplay feature generation tasks and how they supported the respective development activity.

\begin{table*}[t]
\centering
\caption{Prompting strategies used for the evaluated task classes.}
\label{tab:prompting-strategies}
\begin{tabularx}{\textwidth}{l p{0.32\textwidth} X}
\toprule
\textbf{Tasks} & \textbf{Prompting strategy} & \textbf{Purpose} \\
\midrule
R1--R3 &
Self-consistency prompting; data-guided refactoring; structured reasoning prompting &
Generate and compare alternative refactoring variants, use software metrics to guide restructuring, and support a structured rationale for the selected solution. \\
F1--F3 &
Test-guided prompting; few-shot prompting; self-consistency prompting &
Generate feature-specific tests before implementation, align new tests with the existing pytest structure, and compare alternative implementation approaches. \\
\bottomrule
\end{tabularx}
\end{table*}

Evaluation combined quantitative and functional criteria. Quantitative assessment included software metrics such as lines of code (LOC), cyclomatic complexity (CC), cognitive complexity (CGC), code smells (CS), and maintainability index (MI). LOC, CC, CGC, and CS were obtained through SonarQube-based static analysis~\cite{sonarqube_understanding_2026}, while MI was computed with a custom Python metric script based on the standard maintainability-index calculation and applied consistently across all evaluated tasks. In addition to these metrics, the generated or refactored code was validated through unit tests and manual gameplay assessments in order to examine not only structural code quality but also correct behavior within the running game. For the tasks involving gameplay feature generation, this evaluation logic was complemented by feature-specific tests and iterative revision cycles whenever generated solutions failed to meet the expected functionality.

A task was considered successful only if the resulting implementation satisfied the specified functional requirement, passed the relevant unit tests, and behaved correctly in the manual gameplay assessment. A task was considered unsuccessful when repeated refinement no longer produced meaningful progress toward the required functionality, when generated changes increasingly deviated from the original specification, or when an implementation passed code-level tests but still failed during gameplay assessment. These criteria were used to distinguish between correct implementations, implementations that required further prompting, and implementations that remained functionally or semantically insufficient despite additional iterations.

\section{Results}

\begin{table*}[t]
\centering
\caption{Overview of the evaluated refactoring and gameplay feature generation tasks, including iterations and functional outcomes.}
\label{tab:task_overview}
\begin{tabularx}{\textwidth}{lXclX}
\toprule
\textbf{Category} & \textbf{Use case} & \textbf{Iterations} & \textbf{Outcome} & \textbf{Notes} \\
\midrule
Refactoring & State management optimization & 3 & Successful & Unit tests and gameplay assessment passed after correcting pause handling. \\
Refactoring & Redundant asset-loading removal & 7 & Successful & Required corrections for asset compatibility and image scaling. \\
Refactoring & Movement control unification & 5 & Successful & Required corrections to preserve movement-state behavior. \\
Feature generation & Pixel-perfect collision detection & 4 & Successful & Integrated after test- and gameplay-based refinement. \\
Feature generation & Shrink power-up & 8 & Failed & Unit tests passed, but gameplay integration failed due to scaling interactions. \\
Feature generation & Car platform mechanic & 6 & Failed & Unit tests passed, but manual gameplay assessment revealed incorrect landing behavior. \\
\bottomrule
\end{tabularx}
\end{table*}

The results indicate qualitative differences between the selected refactoring tasks and the selected gameplay feature generation tasks in this case study. As shown in Table~\ref{tab:task_overview}, all refactoring tasks were completed successfully in functional terms, whereas only one of the tasks involving gameplay feature generation resulted in a correctly integrated feature. Overall, the observed pattern suggests that GPT-4o handled the selected localized refactoring tasks more reliably than the selected gameplay feature generation tasks in this study.

The iteration counts in Table~\ref{tab:task_overview} should not be interpreted as repeated independent model runs. Rather, they indicate task-specific refinement cycles within a single run, as defined in Section~\ref{sec:methodology}. The reasons for additional iterations differed across tasks: refactoring iterations were mainly caused by missing functionality, compatibility issues, or regressions detected by tests, whereas the failed gameplay feature generation tasks involved deeper integration problems that persisted even after code-level tests had passed. This distinction is important because it suggests that the observed failures were not only caused by syntax errors or missing unit-test coverage, but also by mismatches between generated code and runtime gameplay behavior.

Across the three refactoring tasks, GPT-4o produced functionally stable outcomes and generally improved the structure of the existing code. The three tasks required 3, 7, and 5 iterations, respectively, indicating that useful results could be achieved, but not without task-dependent effort. At the same time, the extent of improvement varied across cases. Quantitatively, the refactoring tasks showed a differentiated pattern: while some structural metrics improved clearly, others changed only marginally or became less favorable, indicating that functional success did not always coincide with uniformly improved metric profiles.

Table~\ref{tab:refactoring_metrics} summarizes selected metric changes for the three refactoring tasks. The table reports lines of code (LOC) as a size measure, cyclomatic complexity (CC) and cognitive complexity (CGC) as complexity measures, code smells (CS) as indicators of potential maintainability issues, and maintainability index (MI) as an aggregate maintainability measure.

\begin{table*}[t]
\centering
\caption{Selected metric changes for the refactoring tasks. Values are reported as before $\rightarrow$ after.}
\label{tab:refactoring_metrics}
\begin{tabularx}{\textwidth}{llcccccX}
\toprule
\textbf{Task} & \textbf{File} & \textbf{LOC} & \textbf{CC} & \textbf{CGC} & \textbf{CS} & \textbf{MI} & \textbf{Interpretation} \\
\midrule
R1 & game.py & 314 $\rightarrow$ 349 & 65 $\rightarrow$ 76 & 118 $\rightarrow$ 96 & 3 $\rightarrow$ 2 & 37.37 $\rightarrow$ 39.52 & Lower cognitive complexity and fewer code smells, but higher LOC and cyclomatic complexity. \\
R2 & assets.py & 118 $\rightarrow$ 45 & 4 $\rightarrow$ 9 & 1 $\rightarrow$ 9 & 1 $\rightarrow$ 1 & 80.83 $\rightarrow$ 88.37 & Strong reduction in file size and improved maintainability index, but higher complexity values. \\
R3 & player.py & 144 $\rightarrow$ 165 & 49 $\rightarrow$ 54 & 59 $\rightarrow$ 54 & 1 $\rightarrow$ 3 & 52.41 $\rightarrow$ 52.04 & Slightly lower cognitive complexity, but more code smells and a marginally lower maintainability index. \\
\bottomrule
\end{tabularx}
\end{table*}

The strongest structural improvement was observed for the refactoring of redundant asset-loading operations, where the affected file became substantially shorter and its maintainability index increased, although its cyclomatic and cognitive complexity also increased. Taken together, the refactoring results suggest that GPT-4o was able to provide practically useful support for this class of modification tasks, even though the quality gains were not equally pronounced in every case.

In the selected tasks involving gameplay feature generation, GPT-4o produced less reliable outcomes. Only the pixel-perfect collision detection feature was implemented successfully, and even this required 4 iterations before the feature behaved as intended. Table~\ref{tab:feature_metrics} summarizes selected metric changes for the gameplay feature generation tasks. For tasks affecting multiple files, LOC, CC, CGC, and CS are aggregated across affected files, while MI is summarized qualitatively because it is file-specific. The table shows that the successfully integrated pixel-perfect collision detection feature required only moderate code changes, whereas the failed feature-generation tasks involved broader changes across files or larger increases in code size and complexity.

\begin{table*}[t]
\centering
\caption{Selected metric changes for the gameplay feature generation tasks. Values are reported as before $\rightarrow$ after.}
\label{tab:feature_metrics}
\begin{tabularx}{\textwidth}{llccccX}
\toprule
\textbf{Task} & \textbf{Files} & \textbf{LOC} & \textbf{CC} & \textbf{CGC} & \textbf{CS} & \textbf{Interpretation} \\
\midrule
F1 & game.py & 314 $\rightarrow$ 318 & 65 $\rightarrow$ 66 & 118 $\rightarrow$ 118 & 3 $\rightarrow$ 3 & Moderate code changes. MI decreased slightly from 37.37 to 36.93, while the feature was successfully integrated. \\
F2 & 6 files & 694 $\rightarrow$ 768 & 138 $\rightarrow$ 140 & 204 $\rightarrow$ 211 & 5 $\rightarrow$ 5 & Broad code changes across several gameplay systems. MI stayed constant or decreased across affected files. The feature still failed during gameplay integration. \\
F3 & 2 files & 458 $\rightarrow$ 487 & 114 $\rightarrow$ 122 & 177 $\rightarrow$ 189 & 4 $\rightarrow$ 5 & Code size and complexity increased in both affected files. MI decreased in both files. Unit tests passed, but gameplay assessment revealed incorrect landing behavior. \\
\bottomrule
\end{tabularx}
\end{table*}

The shrink power-up could not be completed successfully despite 8 refinement attempts, and the car platform mechanic remained problematic as well: although all unit tests passed after 6 iterations, the feature still failed in manual gameplay testing. More precisely, the shrink power-up failed because changes to player-character scaling interfered with existing gameplay mechanics such as jumping, sliding, and weapon behavior, whereas the car platform mechanic failed because the player character still died or moved incorrectly when attempting to land on and move with cars during actual gameplay. These results show that the main difficulties were not limited to superficial code quality issues, but concerned the functional and semantic integration of newly generated gameplay features into the existing game architecture. Within this study, the selected gameplay feature generation tasks exposed more integration-related difficulties than the selected refactoring tasks, but this pattern should be interpreted as exploratory rather than generalizable.

\section{Discussion}

The findings suggest that GPT-4o's behavior in this case study depended strongly on the nature and integration depth of the selected task. In the evaluated refactoring tasks, GPT-4o handled localized code transformations more reliably than the selected tasks involving new gameplay functionality or extensions to existing gameplay systems. This suggests that, in the examined setting, LLM-assisted development was more successful when changes were well-bounded and did not require deep integration into several interdependent gameplay systems. In the refactoring tasks, GPT-4o produced functionally stable outcomes and, in several cases, structural improvements. However, generating new gameplay features required broader integration with existing systems, posing significant challenges. Successful use often required ongoing human input, such as refining prompts, selectively integrating code, and providing corrective feedback.

From a practical perspective, the results suggest that LLMs can be useful as targeted support tools for localized and structurally bounded modification tasks in game development. In such cases, they may help streamline code improvement, exploratory restructuring, and the generation of alternative implementation variants. More complex gameplay features, which require coordination across multiple systems, still demand significant human oversight and validation. The results therefore do not support treating LLM-generated gameplay functionality as immediately production-ready. Instead, they point to the need for careful integration, testing, and developer supervision.

The study also found that code-level validation, such as unit tests, was insufficient to ensure generated gameplay features worked as intended, highlighting the importance of supplementing these with gameplay-oriented evaluation. This was particularly visible in the car platform mechanic, where unit tests passed but manual gameplay assessment still revealed incorrect runtime behavior. For game-development contexts, this distinction is important because functional correctness may depend not only on isolated code behavior, but also on timing, movement, collision handling, animation state, and other runtime interactions that are difficult to capture fully through unit tests alone.

Several limitations should be considered when interpreting these findings. First, the study is based on a single game project, which limits the generalizability of the results to other game genres, architectures, or development contexts. Moreover, the codebase was custom-developed by the authors, which provided full control over the experimental artifact but may also have introduced selection or design bias. Second, only one model, GPT-4o, was evaluated, so the findings cannot be assumed to transfer directly to other LLMs, newer model versions, specialized code models, or agentic coding systems. Third, the study comprised only three refactoring tasks and three tasks involving gameplay feature generation, which is appropriate for a focused exploratory case study but does not allow broader statistical conclusions.

In addition, each task was conducted once rather than repeated across multiple independent model runs, so the study does not account for LLM nondeterminism or run-to-run variation. The tasks were also not pre-calibrated for equivalent difficulty, which means that the observed difference between refactoring and gameplay feature generation cannot be interpreted as a controlled benchmark comparison between the two task classes. Functional success was operationalized through software metrics, unit tests, and manual gameplay assessments, providing a practically relevant evaluation perspective but not fully capturing all aspects of long-term code quality, maintainability, or developer effort.

Finally, the use of LLMs for programming assistance also raises broader concerns regarding responsible use. Iterative prompting workflows consume computational resources, and their use should be balanced against the practical benefits they provide. Generated code may also contain subtle defects despite appearing plausible, which reinforces the need for human oversight, transparent reporting, and validation through both code-level and gameplay-level assessment. In this sense, the study supports a cautious view of LLM-assisted game development: LLMs can provide useful support in bounded tasks, but they should not replace expert review or systematic testing.

\section{Conclusion \& Future Work}

This paper presented an exploratory empirical case study on the use of GPT-4o for refactoring and gameplay feature generation in an existing game software system. Across the selected tasks, GPT-4o handled the refactoring tasks more reliably than the gameplay feature generation tasks, although this pattern should be interpreted in light of the study's exploratory single-case design. The findings suggest that, in the examined setting, LLM-assisted development was more reliable for localized code transformations than for integration-intensive gameplay extensions. At the same time, the study highlights the importance of evaluating generated code not only in structural terms, but also with regard to functional behavior within the running game.

Future work should extend this exploratory approach to additional LLMs, larger and more diverse task sets, repeated independent model runs, and different game projects in order to assess whether the observed pattern remains stable across broader development contexts. Such studies should also consider task-difficulty calibration before comparing different classes of development tasks and examine whether emerging agentic coding systems can reduce the amount of human guidance required in iterative workflows of this kind.

\printbibliography

@inproceedings{ross_programmers_2023,
	title = {The {Programmer}’s {Assistant}: {Conversational} {Interaction} with a {Large} {Language} {Model} for {Software} {Development}},
	isbn = {979-8-4007-0106-1},
	shorttitle = {The {Programmer}’s {Assistant}},
	url = {https://dl.acm.org/doi/10.1145/3581641.3584037},
	doi = {10.1145/3581641.3584037},
	urldate = {2026-03-12},
	booktitle = {Proc. 28th Int. Conf. Intelligent User Interfaces},
	publisher = {ACM},
	author = {Ross, Steven I. and Martinez, Fernando and Houde, Stephanie and Muller, Michael and Weisz, Justin D.},
	month = mar,
	year = {2023},
	pages = {491--514},
}

@article{nejjar_llms_2025,
	title = {{LLMs} for science: {Usage} for code generation and data analysis},
	volume = {37},
	issn = {2047-7473, 2047-7481},
	shorttitle = {{LLMs} for science},
	url = {https://onlinelibrary.wiley.com/doi/10.1002/smr.2723},
	doi = {10.1002/smr.2723},
	number = {1},
	urldate = {2026-03-12},
	journal = {Journal of Software: Evolution and Process},
	author = {Nejjar, Mohamed and Zacharias, Luca and Stiehle, Fabian and Weber, Ingo},
	month = jan,
	year = {2025},
	pages = {e2723},
}

@inproceedings{liu_guiding_2024,
	title = {Guiding {ChatGPT} for {Better} {Code} {Generation}: {An} {Empirical} {Study}},
	copyright = {https://doi.org/10.15223/policy-029},
	isbn = {979-8-3503-3066-3},
	shorttitle = {Guiding {ChatGPT} for {Better} {Code} {Generation}},
	url = {https://ieeexplore.ieee.org/document/10589825/},
	doi = {10.1109/SANER60148.2024.00018},
	urldate = {2026-03-12},
	booktitle = {IEEE Int. Conf. Software Analysis, Evolution and Reengineering},
	publisher = {IEEE},
	author = {Liu, Chao and Bao, Xuanlin and Zhang, Hongyu and Zhang, Neng and Hu, Haibo and Zhang, Xiaohong and Yan, Meng},
	month = mar,
	year = {2024},
	pages = {102--113},
}

@article{gallotta_large_2024,
	title = {Large {Language} {Models} and {Games}: {A} {Survey} and {Roadmap}},
	issn = {2475-1502, 2475-1510},
	shorttitle = {Large {Language} {Models} and {Games}},
	url = {https://ieeexplore.ieee.org/document/10680313/},
	doi = {10.1109/TG.2024.3461510},
	journal = {IEEE Trans. Games},
	author = {Gallotta, Roberto and Todd, Graham and Zammit, Marvin and Earle, Sam and Liapis, Antonios and Togelius, Julian and Yannakakis, Georgios N.},
	year = {2024},
	pages = {1--18},
}

@article{gu_effectiveness_2025,
	title = {On the {Effectiveness} of {Large} {Language} {Models} in {Domain}-{Specific} {Code} {Generation}},
	volume = {34},
	issn = {1049-331X, 1557-7392},
	url = {https://dl.acm.org/doi/10.1145/3697012},
	doi = {10.1145/3697012},
	number = {3},
	urldate = {2026-03-12},
	journal = {ACM Transactions on Software Engineering and Methodology},
	author = {Gu, Xiaodong and Chen, Meng and Lin, Yalan and Hu, Yuhan and Zhang, Hongyu and Wan, Chengcheng and Wei, Zhao and Xu, Yong and Wang, Juhong},
	month = mar,
	year = {2025},
	pages = {1--22},
}

@inproceedings{marini_leveraging_2024,
	title = {Leveraging {Large} {Language} {Models} for {Efficient} {Failure} {Analysis} in {Game} {Development}},
	copyright = {https://doi.org/10.15223/policy-029},
	isbn = {979-8-3503-5067-8},
	url = {https://ieeexplore.ieee.org/document/10645540/},
	doi = {10.1109/CoG60054.2024.10645540},
	urldate = {2026-03-12},
	booktitle = {IEEE Conf. Games (CoG)},
	publisher = {IEEE},
	author = {Marini, Leonardo and Gisslén, Linus and Sestini, Alessandro},
	month = aug,
	year = {2024},
	pages = {1--8},
}

@article{liu_no_2024,
	title = {No {Need} to {Lift} a {Finger} {Anymore}? {Assessing} the {Quality} of {Code} {Generation} by {ChatGPT}},
	volume = {50},
	copyright = {https://ieeexplore.ieee.org/Xplorehelp/downloads/license-information/IEEE.html},
	issn = {0098-5589, 1939-3520, 2326-3881},
	shorttitle = {No {Need} to {Lift} a {Finger} {Anymore}?},
	url = {https://ieeexplore.ieee.org/document/10507163/},
	doi = {10.1109/TSE.2024.3392499},
	number = {6},
	urldate = {2026-03-12},
	journal = {IEEE Transactions on Software Engineering},
	author = {Liu, Zhijie and Tang, Yutian and Luo, Xiapu and Zhou, Yuming and Zhang, Liang Feng},
	month = jun,
	year = {2024},
	pages = {1548--1584},
}

@article{sagodi_methodology_2024,
	title = {Methodology for {Code} {Synthesis} {Evaluation} of {LLMs} {Presented} by a {Case} {Study} of {ChatGPT} and {Copilot}},
	volume = {12},
	copyright = {https://creativecommons.org/licenses/by-nc-nd/4.0/},
	issn = {2169-3536},
	url = {https://ieeexplore.ieee.org/document/10535504/},
	doi = {10.1109/ACCESS.2024.3403858},
	urldate = {2026-03-12},
	journal = {IEEE Access},
	author = {Ságodi, Zoltán and Siket, István and Ferenc, Rudolf},
	year = {2024},
	pages = {72303--72316},
}

@article{claypool_latency_2006,
	title = {Latency and player actions in online games},
	volume = {49},
	issn = {0001-0782, 1557-7317},
	url = {https://dl.acm.org/doi/10.1145/1167838.1167860},
	doi = {10.1145/1167838.1167860},
	number = {11},
	urldate = {2026-03-12},
	journal = {Commun. ACM},
	author = {Claypool, Mark and Claypool, Kajal},
	month = nov,
	year = {2006},
	pages = {40--45},
}

@techreport{openaiGPT4oSystemCard2024,
  title = {{{GPT-4o System Card}}},
  author = {OpenAI},
  year = 2024,
  url = {https://openai.com/index/gpt-4o-system-card/},
  urldate = {2025-03-27},
  langid = {american},
  institution = {OpenAI},
  note = {\url{https://openai.com/index/gpt-4o-system-card/}}
}

@inproceedings{jin_can_2024,
	title = {Can {ChatGPT} {Support} {Developers}? {An} {Empirical} {Evaluation} of {Large} {Language} {Models} for {Code} {Generation}},
	isbn = {979-8-4007-0587-8},
	shorttitle = {Can {ChatGPT} {Support} {Developers}?},
	url = {https://dl.acm.org/doi/10.1145/3643991.3645074},
	doi = {10.1145/3643991.3645074},
	urldate = {2026-03-12},
	booktitle = {Proc. 21st Int. Conf. Mining Software Repositories},
	publisher = {ACM},
	author = {Jin, Kailun and Wang, Chung-Yu and Pham, Hung Viet and Hemmati, Hadi},
	month = apr,
	year = {2024},
	pages = {167--171},
}

@article{liu_refining_2024,
	title = {Refining {ChatGPT}-{Generated} {Code}: {Characterizing} and {Mitigating} {Code} {Quality} {Issues}},
	volume = {33},
	issn = {1049-331X, 1557-7392},
	shorttitle = {Refining {ChatGPT}-{Generated} {Code}},
	url = {https://dl.acm.org/doi/10.1145/3643674},
	doi = {10.1145/3643674},
	number = {5},
	urldate = {2026-03-12},
	journal = {ACM Transactions on Software Engineering and Methodology},
	author = {Liu, Yue and Le-Cong, Thanh and Widyasari, Ratnadira and Tantithamthavorn, Chakkrit and Li, Li and Le, Xuan-Bach D. and Lo, David},
	month = jun,
	year = {2024},
	pages = {1--26},
}

@misc{sweetser_large_2024,
	title = {Large {Language} {Models} and {Video} {Games}: {A} {Preliminary} {Scoping} {Review}},
	copyright = {arXiv.org perpetual, non-exclusive license},
	shorttitle = {Large {Language} {Models} and {Video} {Games}},
	url = {https://arxiv.org/abs/2403.02613},
	doi = {10.48550/ARXIV.2403.02613},
	publisher = {arXiv},
	author = {Sweetser, Penny},
	year = {2024},
}

@article{yang_gpt_2025_IEEE_TOG,
  author={Yang, Daijin and Kleinman, Erica and Harteveld, Casper},
  journal={IEEE Transactions on Games}, 
  title={GPT for Games: An Updated Scoping Review (2020-2024)}, 
  year={2025},
  volume={},
  number={},
  pages={1-16},
  doi={10.1109/TG.2025.3563780}
}

@inproceedings{shirafuji_refactoring_2023,
	title = {Refactoring {Programs} {Using} {Large} {Language} {Models} with {Few}-{Shot} {Examples}},
	copyright = {https://doi.org/10.15223/policy-029},
	isbn = {979-8-3503-4417-2},
	url = {https://ieeexplore.ieee.org/document/10479398/},
	doi = {10.1109/APSEC60848.2023.00025},
	urldate = {2026-03-12},
	booktitle = {30th Asia-Pacific Softw. Eng. Conf. (APSEC)},
	publisher = {IEEE},
	author = {Shirafuji, Atsushi and Oda, Yusuke and Suzuki, Jun and Morishita, Makoto and Watanobe, Yutaka},
	month = dec,
	year = {2023},
	pages = {151--160},
}

@misc{liu_empirical_2024,
	title = {An {Empirical} {Study} on the {Potential} of {LLMs} in {Automated} {Software} {Refactoring}},
	url = {https://arxiv.org/abs/2411.04444},
	doi = {10.48550/ARXIV.2411.04444},
	publisher = {arXiv},
	author = {Liu, Bo and Jiang, Yanjie and Zhang, Yuxia and Niu, Nan and Li, Guangjie and Liu, Hui},
	year = {2024},
}

@inproceedings{siddiq_quality_2024,
	title = {Quality {Assessment} of {ChatGPT} {Generated} {Code} and their {Use} by {Developers}},
	isbn = {979-8-4007-0587-8},
	url = {https://dl.acm.org/doi/10.1145/3643991.3645071},
	doi = {10.1145/3643991.3645071},
	urldate = {2026-03-12},
	booktitle = {Proceedings of the 21st {International} {Conference} on {Mining} {Software} {Repositories}},
	publisher = {ACM},
	author = {Siddiq, Mohammed Latif and Roney, Lindsay and Zhang, Jiahao and Santos, Joanna Cecilia Da Silva},
	month = apr,
	year = {2024},
	pages = {152--156},
}

@misc{rasnayaka_empirical_2024,
	title = {An {Empirical} {Study} on {Usage} and {Perceptions} of {LLMs} in a {Software} {Engineering} {Project}},
	url = {https://arxiv.org/abs/2401.16186},
	doi = {10.48550/ARXIV.2401.16186},
	publisher = {arXiv},
	author = {Rasnayaka, Sanka and Wang, Guanlin and Shariffdeen, Ridwan and Iyer, Ganesh Neelakanta},
	year = {2024},
}

@misc{sahoo_systematic_2024,
	title = {A {Systematic} {Survey} of {Prompt} {Engineering} in {Large} {Language} {Models}: {Techniques} and {Applications}},
	copyright = {Creative Commons Attribution 4.0 International},
	shorttitle = {A {Systematic} {Survey} of {Prompt} {Engineering} in {Large} {Language} {Models}},
	url = {https://arxiv.org/abs/2402.07927},
	doi = {10.48550/ARXIV.2402.07927},
	publisher = {arXiv},
	author = {Sahoo, Pranab and Singh, Ayush Kumar and Saha, Sriparna and Jain, Vinija and Mondal, Samrat and Chadha, Aman},
	year = {2024},
}

@article{knoth_ai_2024,
	title = {{AI} literacy and its implications for prompt engineering strategies},
	volume = {6},
	issn = {2666920X},
	url = {https://linkinghub.elsevier.com/retrieve/pii/S2666920X24000262},
	doi = {10.1016/j.caeai.2024.100225},
	urldate = {2026-03-12},
	journal = {Comput. Educ.: Artif. Intell.},
	author = {Knoth, Nils and Tolzin, Antonia and Janson, Andreas and Leimeister, Jan Marco},
	month = jun,
	year = {2024},
	pages = {100225},
}

@inproceedings{mathews_test-driven_2024,
	title = {Test-{Driven} {Development} and {LLM}-based {Code} {Generation}},
	isbn = {979-8-4007-1248-7},
	url = {https://dl.acm.org/doi/10.1145/3691620.3695527},
	doi = {10.1145/3691620.3695527},
	urldate = {2026-03-12},
	booktitle = {Proc. 39th IEEE/ACM Int. Conf. Automated Software Engineering},
	publisher = {ACM},
	author = {Mathews, Noble Saji and Nagappan, Meiyappan},
	month = oct,
	year = {2024},
	pages = {1583--1594},
}

@misc{cordeiro_empirical_2024,
	title = {An {Empirical} {Study} on the {Code} {Refactoring} {Capability} of {Large} {Language} {Models}},
	copyright = {Creative Commons Attribution 4.0 International},
	url = {https://arxiv.org/abs/2411.02320},
	doi = {10.48550/ARXIV.2411.02320},
	urldate = {2026-03-12},
	publisher = {arXiv},
	author = {Cordeiro, Jonathan and Noei, Shayan and Zou, Ying},
	year = {2024},
}

@misc{sonarqube_understanding_2026,
	title = {Understanding measures and metrics {\textbar} {SonarQube} {Server} {\textbar} {Sonar} {Documentation}},
	url = {https://docs.sonarsource.com/sonarqube-server/user-guide/code-metrics/metrics-definition},
	urldate = {2026-05-27},
	author = {SonarQube},
	year = {2026},
    note = {\url{https://docs.sonarsource.com/sonarqube-server/user-guide/code-metrics/metrics-definition}}
}

\end{document}